\documentclass[preprint]{elsarticle}

\usepackage{lineno,hyperref}
\usepackage{amsmath,amssymb}
\usepackage{graphicx}

\usepackage[english]{babel}
\usepackage[T2A]{fontenc}

\usepackage{comment} 

\modulolinenumbers[10]
\modulolinenumbers[1]

\begin{document}
\begin{frontmatter}
\title{New type of chimera structures in a ring of bistable FitzHugh-Nagumo oscillators with nonlocal interaction}

\author{I.A. Shepelev}
\ead{igor\_sar@li.ru}
\address{Department of Physics, Saratov National Research State University,83 Astrakhanskaya Street, Saratov, 410012, Russia} 
\author{T.E. Vadivasova}
\ead{vadivasovate@yandex.ru}
\address{Department of Physics, Saratov National Research State University,83 Astrakhanskaya Street, Saratov, 410012, Russia} 
\author{G.I. Strelkova}
\ead{strelkovagi@info.sgu.ru}
\address{Department of Physics, Saratov National Research State University,83 Astrakhanskaya Street, Saratov, 410012, Russia} 
\author{V.S. Anishchenko}
\ead{wadim@info.sgu.ru}
\address{Department of Physics, Saratov National Research State University,83 Astrakhanskaya Street, Saratov, 410012, Russia}

\begin{abstract}

We study the spatiotemporal dynamics of a ring of nonlocally coupled FitzHugh-Nagumo oscillators in the bistable regime. A new type of chimera patterns has been found in the noise-free network and when isolated elements do not oscillate. The region of existence of these structures has been explored when the coupling range and the coupling strength between the network elements are varied.

\end{abstract}

\end{frontmatter}

\maketitle

\section*{Introduction}
Recently, chimera states, which are characterized by the alternation of clusters with coherent and incoherent behavior of interacting elements, were found in a variety of oscillatory ensembles with nonlocal coupling (see, for example, the review \cite{Panaggio-2015}). The elements of such ensembles can be oscillators with significantly different behavior, e.g. phase oscillators  \cite{Kuramoto-2002, Abrams-2004, Abrams-2006, Omel'chenko-2012, Maistrenko-2015}, self-sustained oscillatory systems with periodic dynamics \cite{Tinsley-2012, Zakharova-2014, Kapitaniak-2014, Omelchenko-2015a, Vullings-2016}, chaotic oscillators and chaotic return maps \cite{Omelchenko-2011, Omelchenko-2012, Semenova-2015, Anishchenko-2016, Vadivasova-2016}. The FitzHugh-Nagumo (FHN) oscillator is not an exception in this respect. Chimera structures have been observed in a network of FHN oscillators both in the case of self-sustained dynamics of elements \cite{Omelchenko-2013, Omelchenko-2015b} and in the excitable regime \cite{Semenova-2016}. In the last case, the appearance of chimera states results from the influence of uncorrelated noise with a certain intensity on the network elements.

The FHN oscillator was proposed in \cite{FitzHugh-1961, Nagumo-1962} as a simplified model of neuron dynamics. Since then it is widely used as a classical example of an excitable system. The dynamics of ensembles of FHN oscillators is of great interest which is related to simulations of the processes in neural fibres and muscular tissues (see, for example, \cite{Ermentrout-2007, Gorelova-1983, Huang-2004, Lancaster-2010}). At the same time, the excitable regime is not the only type of the FHN oscillator behavior. For a special equation form and depending on the parameter values, this oscillator can also demonstrate self-sustained oscillations or can operate in the bistable regime with two stable equilibrium points. If FHN oscillators are diffusively coupled in a ring, then the emergence of traveling waves can be observed in all the three regimes of the isolated element dynamics. The peculiarities of traveling waves are very similar to the cases of excitable and bistable regimes \cite{Shepelev-2016}.
  
In the present paper we study the effect of nonlocal interaction on the spatiotemporal dynamics of a ring of FHN oscillators in the bistable regime. We note that each isolated (uncoupled) element of the ring has two stable equilibrium points and does not oscillate.  Our numerical simulations have shown that not only traveling waves can emerge for nonlocal coupling. We have also found a special type of chimera patterns which are different from the previously studied ones. Although the ring elements are not self-sustained oscillators, the observed chimera structures can exist without random or other external forces. 

\section{System under study}

The system under study is the ring of FHN oscillators, which is described by the following ordinary differential equations:
\begin{equation}
\begin{array}{l}
\varepsilon \dot{x}_i = x_i - \dfrac{x_i^3}3 -y_i + \dfrac \sigma {2P}
\sum\limits_{\tiny k=i-P}^{\tiny i+P}
\left(x_k - x_i \right),\\
\dot y_i = \gamma x_i - y_i - \beta ,
\\ [6pt]
x_{i+N}(t) = x_{i}(t),~~y_{i+N}(t) = y_{i}(t)~~i=1,...N.
\end{array}
\label{eq:main}
\end{equation}
Here $x_{i},~y_{i}$ are dynamical variables defining the state of the $i$th oscillator, 
$i$ is the oscillator number in the ring (discrete spatial coordinate), $t$ is the time variable, $\varepsilon$, $\gamma$, and $\beta$ are the parameters having the same values for all the oscillators. The whole number of elements in the ensemble is equal to $N=300$. The interaction has a nonlocal character. Each element of the ensemble is coupled with $P$ neighbors on both sides. The coupling strength is controlled by the parameter $\sigma$. The quantity $r=P/N$ is usually called as the coupling range.

We now consider how the behavior of a single FHN oscillator changes when varying its parameters. We fix $\varepsilon=0.2$ and $\beta=0.001$. The parameter $\beta$ is chosen to be small in order to realize the bistable regime near the threshold of self-sustained oscillations ($0.001 \leq \beta \leq 0.002$). At the same time, a small deviation of $\beta$ from nonzero value excludes the full symmetry of equilibria and related effects \cite{Shepelev-2015}. For the chosen values of $\beta$ and $\gamma < 0.718$, a single oscillator is in the bistable regime with two stable equilibrium points (foci). When increasing $\gamma$ the transition to self-sustained oscillations takes place. The dissipation in the oscillator increases with decreasing $\gamma$ and the stable foci transform to stable nodes.

\begin{figure}[htbp]
\centering
\includegraphics[width=0.6\linewidth]{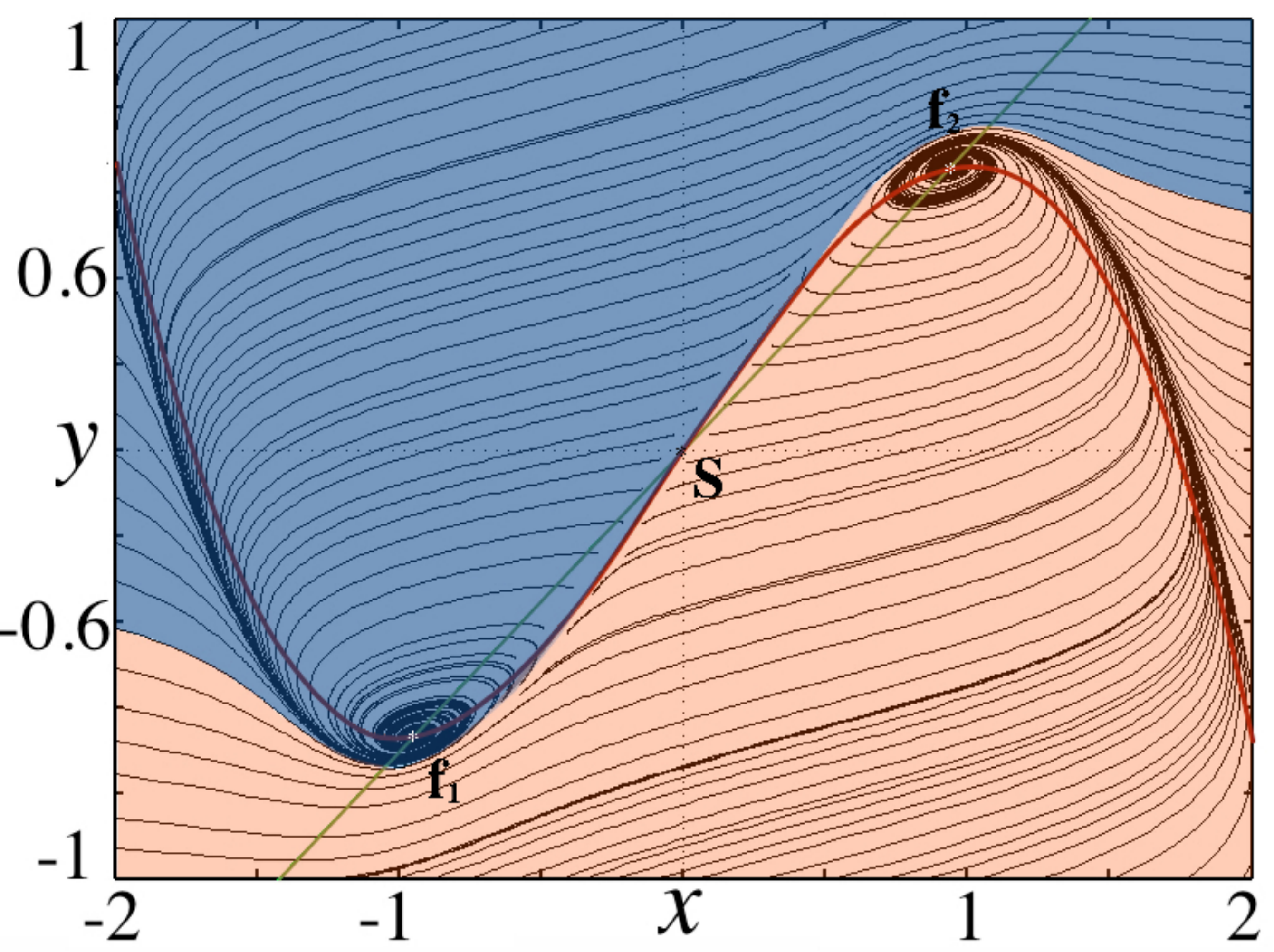}
\caption{Characteristic trajectories and basins of attraction of two stable equilibrium points in a single FHN oscillator. The red line is the nullcline $\dot{x}=0$ and the green line is the nullcline $\dot{y}=0$. The basins of attraction of two foci are highlighted by two tones: light (pink) online) and dark (blue). Parameters: $\varepsilon=0.2$, $\gamma=0.7$, and $\beta=0.001$. (Color online).
}
\label{pic:nullklines}
\end{figure}

The phase portrait of the bistable FHN oscillator is presented in Fig.~\ref{pic:nullklines}. Characteristic trajectories and nullclines intercrossing in three equilibrium points (two stable foci and a saddle) are shown in the phase plane. Basins of attractions of two stable equilibria are marked by two tones (colors online).
%
%
%

It should be noted that the system under study Eq.\eqref{eq:main} is different from the models of the ring of FHN oscillators where chimera regimes were observed earlier.

A simple dissipative coupling for only one variable in the first equation of the oscillators is used in Eq.\eqref{eq:main} in contrast to the rings considered in \cite{Omelchenko-2013, Omelchenko-2015b, Semenova-2016}, where a special kind of coupling was used. This coupling provides phase shifts near to $\pi/2$ between interacting oscillators. Noise sources are not included in Eq.\eqref{eq:main} as compared to the model considered in \cite{Semenova-2016}. An essential feature of the ensemble Eq.\eqref{eq:main} is the bistable dynamics of the oscillators while in the previous works the oscillators were either in the self-sustained regime \cite{Omelchenko-2013, Omelchenko-2015b} or in the excitable regime \cite{Semenova-2016}.

\section{Diagram of regimes in the ($r$,$\sigma$) parameter plane. Description of the basic regimes}

\begin{figure}[htbp]
\centering
\includegraphics[width=0.65\linewidth]{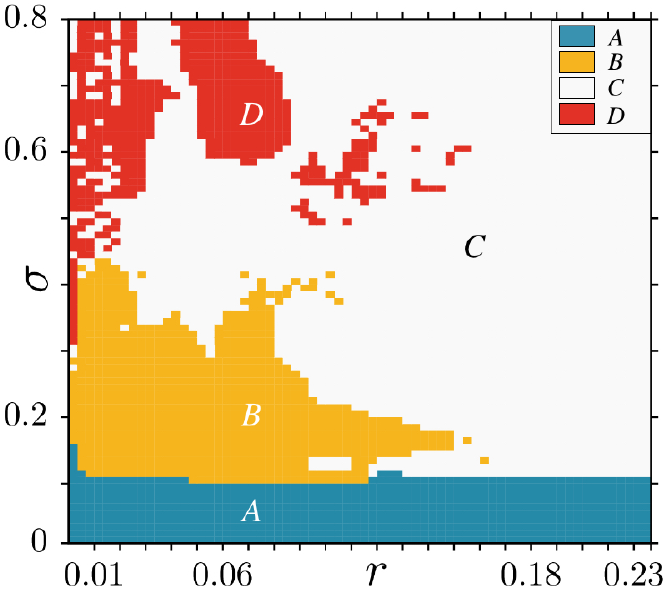}
\caption{Diagram of regimes for the system Eq.\eqref{eq:main} in the ($r$,$\sigma$) parameter plane. $A$: irregular stationary structures; $B$: chimera regime; $C$: spatio-uniform equilibrium regime, $D$: traveling waves. Parameters: $\varepsilon=0.2$, $\gamma=0.7$, $\beta=0.001$ $N$=300}
\label{pic:diagram}
\end{figure}

A diagram of regimes for the network Eq.\eqref{eq:main} is plotted in the ($r$,$\sigma$) parameter plane for the fixed $\varepsilon=0.2$, $\gamma=0.7$, and $\beta=0.001$. That is done to better understand  the effect of nonlocal coupling on the behavior of the ring Eq.\eqref{eq:main} with the bistable dynamics of the elements. This diagram is shown in Fig.~\ref{pic:diagram}. The same realization of random values was used as initial conditions for plotting the diagram. These values were produced by a random number generator with the uniform probability distribution in the unit interval $x_{i}(t_{0}) \in[-1;~1];~y_{i}(t_{0})\in[-1;~1],~ i=1,2,...,N$. Thus, at the initial time $t_{0}$ the oscillators are distributed between two potential wells (left and right) corresponding to basins of attractions of two foci in a single oscillator (see Fig.~\ref{pic:nullklines}). If the initial conditions of all the oscillators are chosen within only one well, then a spatio-uniform equilibrium regime, when all oscillators rest in one equilibrium, is set in the ring with time.

%
%

\begin{figure}[!ht]
\centering
\parbox[c]{.4\linewidth}{
  \includegraphics[width=\linewidth]{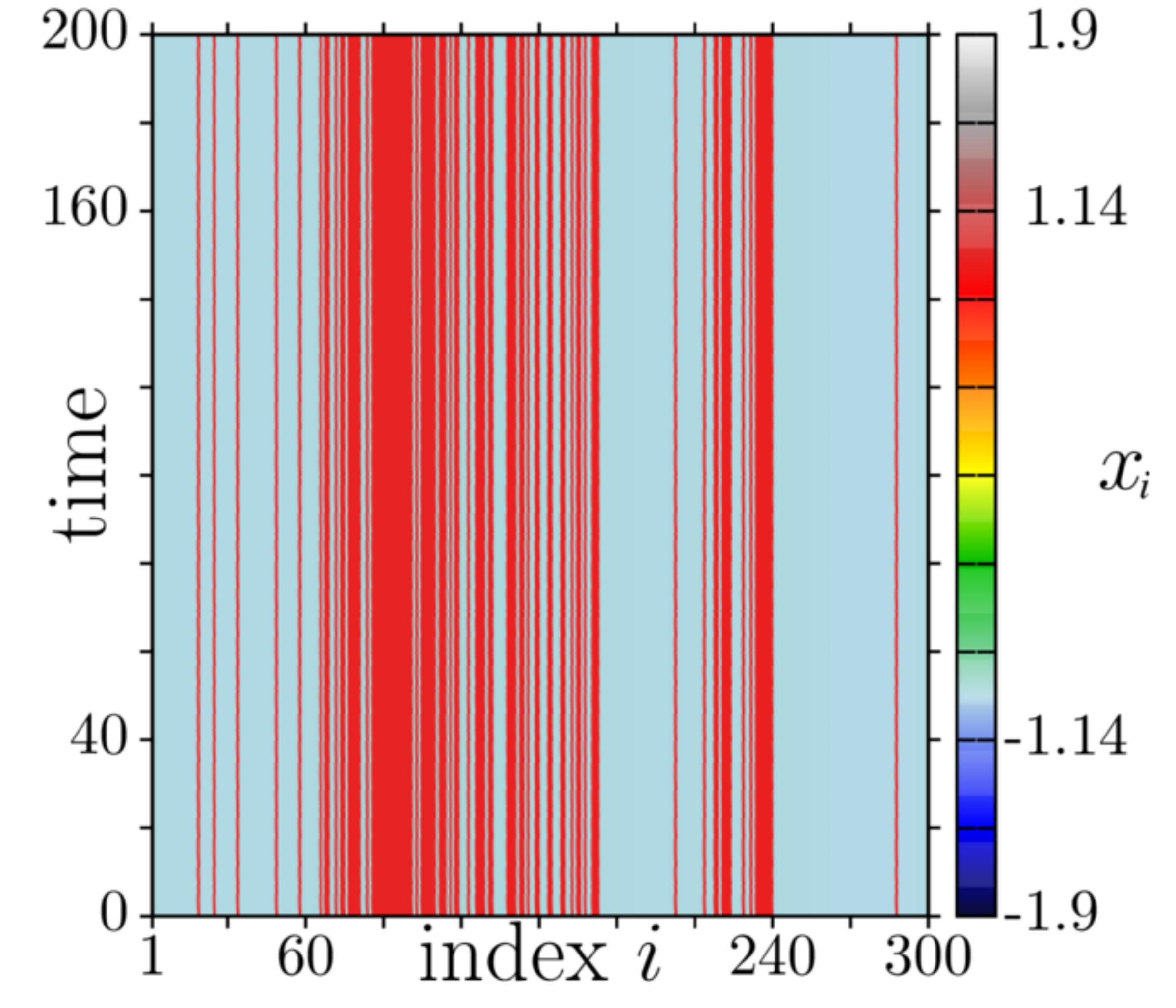}
\center (a)
\includegraphics[width=\linewidth]{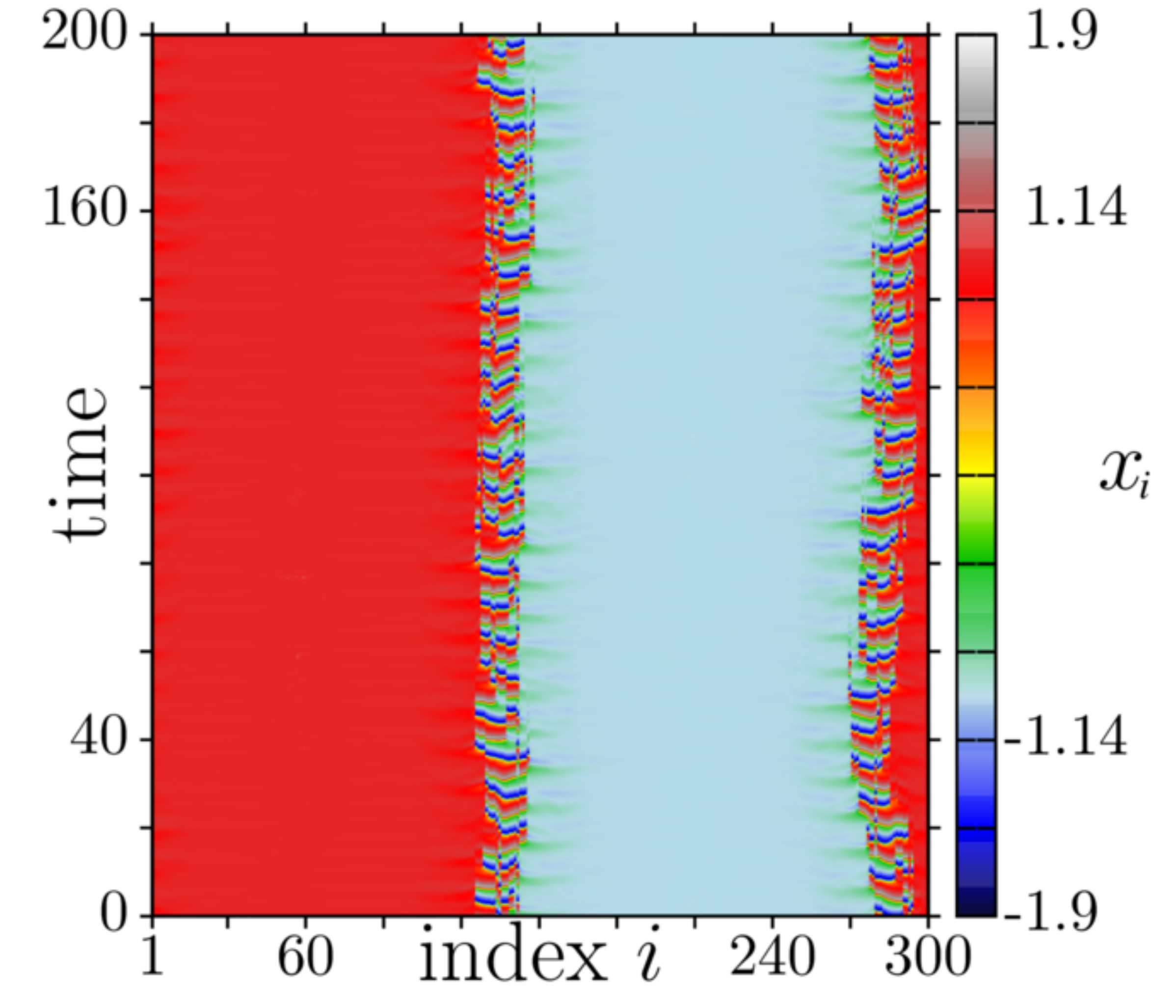}
\center (c)
}
\parbox[c]{.4\linewidth}{
  \includegraphics[width=\linewidth]{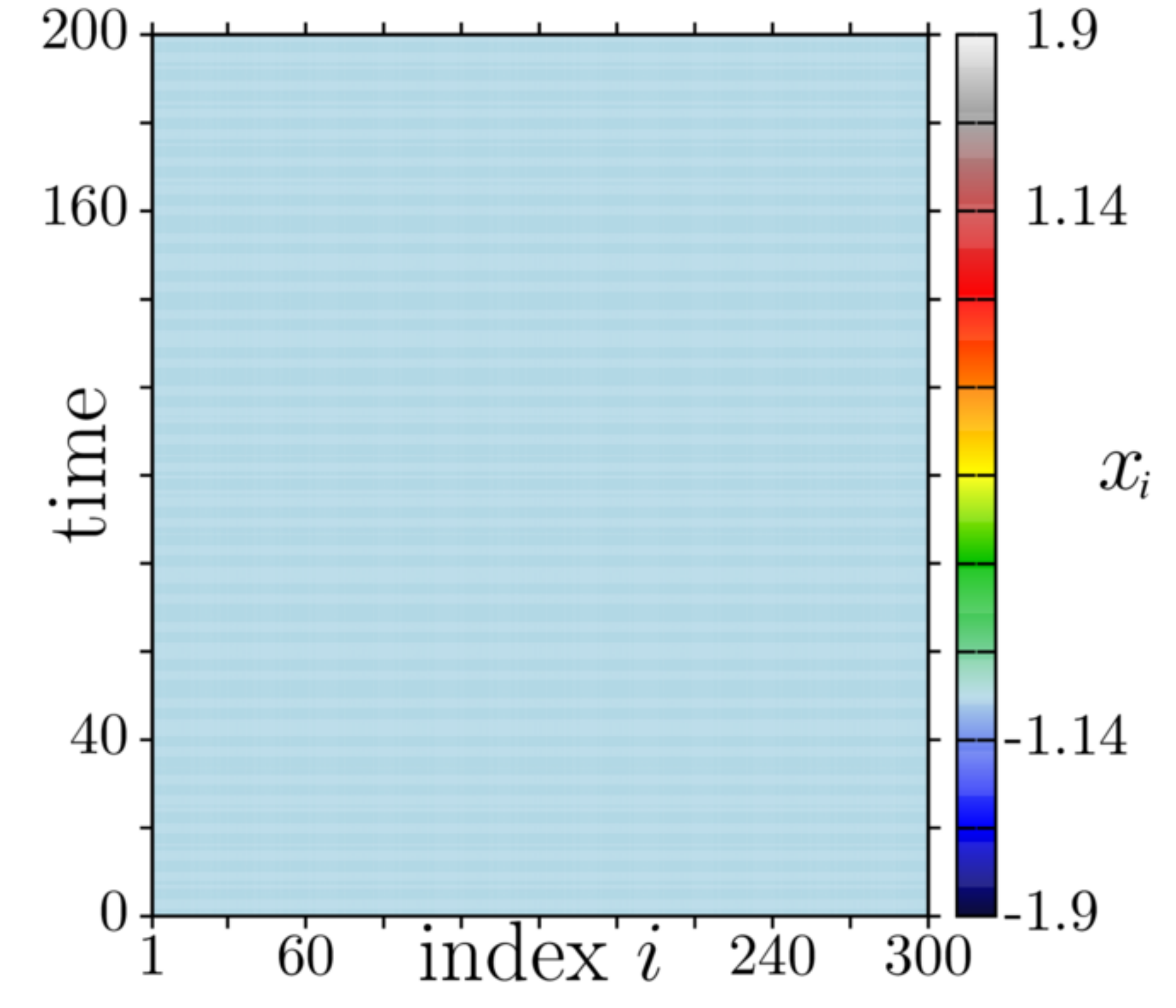}
\center (b)
  \includegraphics[width=\linewidth]{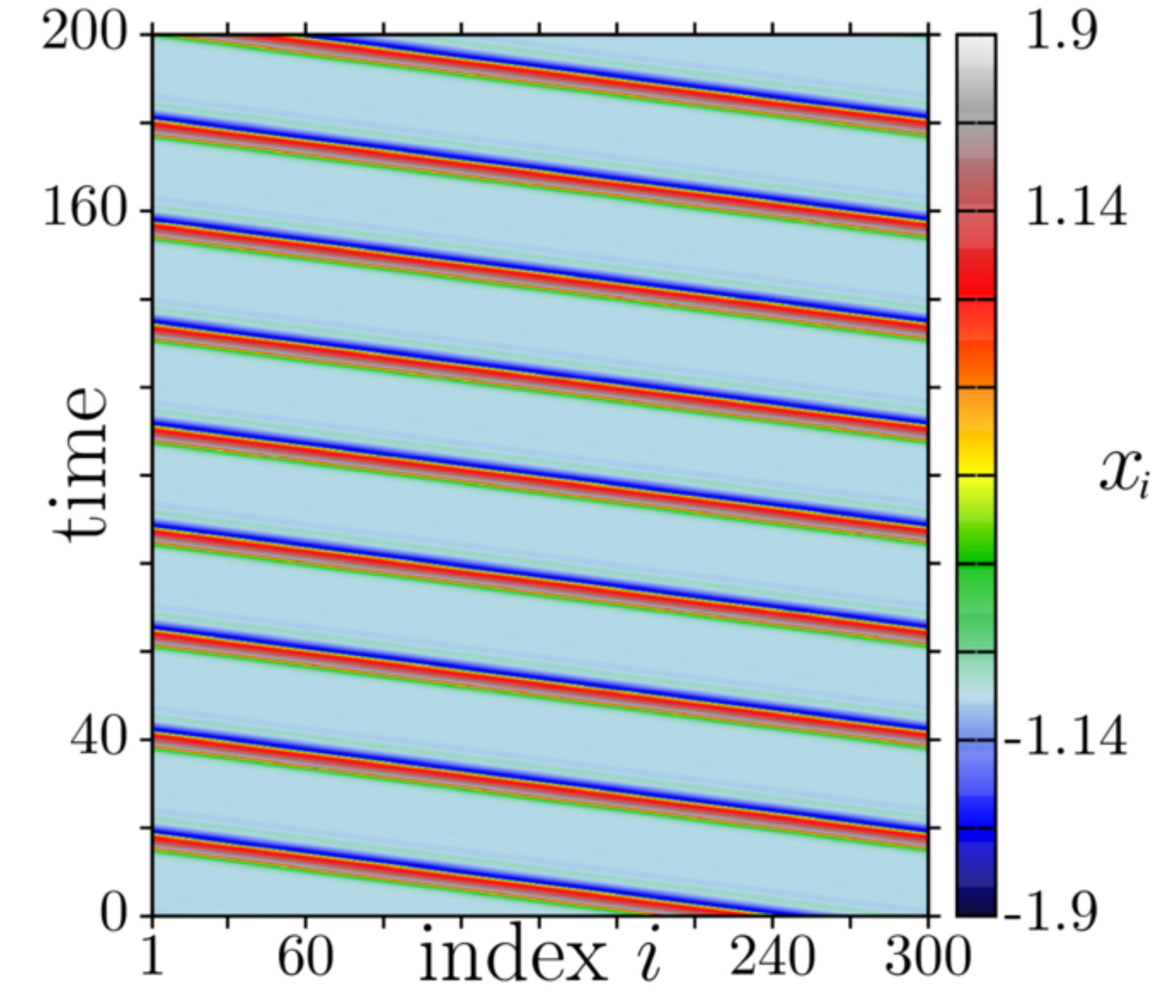}
\center (d)
}

\caption{Basic regimes in the ring Eq.\eqref{eq:main} for different values $\sigma$. (a): irregular stationary structure for $\sigma=0.07$, (b): spatio-uniform equilibrium regime for $\sigma=0.43$, (c): chimera state for $\sigma=0.34$, (d): traveling wave for $\sigma=0.68$. The regimes are illustrated by space-time plots for variable $x_i$. Parameters: $r=0.077$, $\varepsilon=0.2$, $\gamma=0.7$, $\beta=0.001$, $N$=300. (Colors online)}
\label{pic:regimes}
\end{figure}

Different tones (colors online) in Fig.~\ref{pic:diagram} correspond to four basic regimes, which are typical for the system Eq.\eqref{eq:main} in case of the bistable dynamics of the elements. 
For convenience, these domains are also marked by letters $A,~B,~C~$, and $~D$ in the diagram. The regime of irregular stationary structures (region $A$ highlighted by the blue color online) is observed for relatively small values of the coupling strength $\sigma$ and independently of the coupling range $r$. The interval of $\sigma$ values, where such patterns are observed, remains practically the same for any value of $r$. This regime is shown in Fig.~\ref{pic:regimes}(a). A specific form of the spatial structure is defined by the initial distribution of the oscillators between the potential wells and the value of $\sigma$. Increasing the coupling strength provides the growth of clusters corresponding to the distribution of adjacent oscillators in one well, and the spatial structure is simplified. Similar spatial patterns are well-known for the chains and lattices of bistable elements with local dissipative coupling (see, for example, \cite{Nekorkin-2002}). So, their existence for weak nonlocal coupling is well expected.

%
As the coupling increases, two scenarios can be observed. The first one is the transition to a spatio-uniform stationary regime (grey area $C$ in Fig.~\ref{pic:diagram}), when all the oscillators rest in one equilibrium state (left or right) and there are no oscillations in the system Eq.\eqref{eq:main}. The second way is the transition to a regime with the spatial structure corresponding to a chimera state \cite{Panaggio-2015, Kuramoto-2002, Abrams-2004} (yellow domain $C$ online). The spatio-uniform equilibrium regime is shown in Fig.~\ref{pic:regimes}(b). The chimera structure is exemplified in Fig.~\ref{pic:regimes}(c). 
This structure consists of two coherence and two incoherence clusters. The coherence clusters correspond to wide strips of uniform light and dark tones (red and blue colors online) in the space-time plot shown in Fig.~\ref{pic:regimes}(c)). The incoherence clusters look like as narrow strips with an irregular alternation of light and dark tones (red and blue colors online). This chimera will be considered in details in the next section of this article. It should be noted that the chimera regime is observed only for relatively small values of the coupling range $r<0.18$ and can exist up to the transition to local interaction for $r=0.00333...$.

The regime of traveling waves in region $D$ (Fig.~\ref{pic:diagram}) (red online) is shown in Fig.~\ref{pic:regimes}(d). The traveling waves are observed for relatively small values of $r$, as well as chimera structures, but for larger values of $\sigma$. Periodic traveling waves propagate in the ring in any of the two directions. The phase velocity of these waves increases with the growth of the coupling strength. It should be noted that for chosen initial conditions, only one-wave mode (only one full wavelength along the ring) is realized in the entire area of existence of traveling waves.

\section{Peculiarities of chimera structures in the ring of bistable FHN oscillators}

As mentioned in the recent works \cite{Omelchenko-2013, Omelchenko-2015b, Semenova-2016}, chimera states were found in a ring of FHN oscillators in the self-sustained and excitable regimes. The first case corresponds to chimera states which are analogue to the phase chimera in ensembles of Kuramoto phase oscillators with nonlocal coupling. Noise-induced chimeras were observed in the second case. In both cases, coupling between oscillators was introduced by the same complicated method. If we choose a simpler coupling type, which is used in the model Eq.\eqref{eq:main}, then chimera states cannot be observed in the self-sustained and excitable regimes. The peculiarity of the chimera regime which has been found in our studies is its existence only for small values of the coupling range $r$.

The chimera in the ring of bistable FHN oscillators Eq.\eqref{eq:main} has the following specific features: the oscillators from the coherence cluster are located in the bottom of the same potential well and are almost motionless. There are at least two coherence clusters corresponding to two equilibrium states (two wells). The oscillators located in the boundary region form two incoherence clusters. These oscillators are randomly distributed between the wells at one time moment and, moreover, they produce incoherent oscillations in time including switchings between the two wells. The spatial location of these domains, where irregular oscillations are observed, remains practically unchanged in time.

\begin{figure}[!ht]
\centering
\parbox[c]{.38\linewidth}{
  \includegraphics[width=\linewidth]{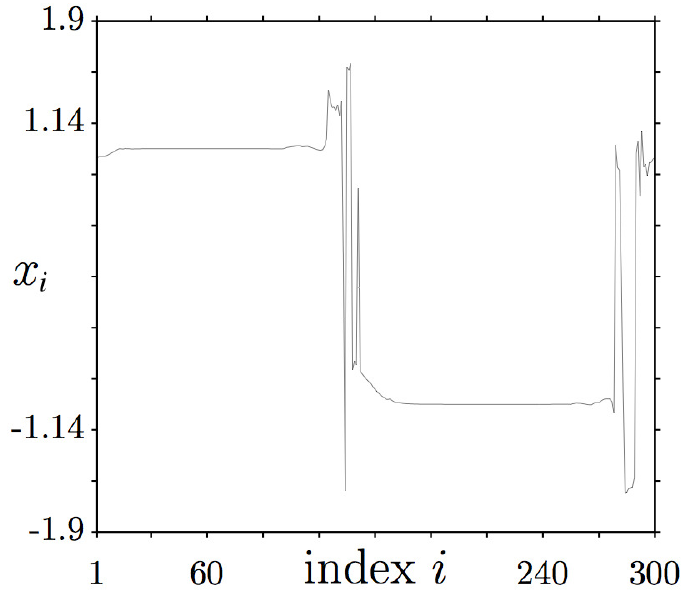}
\center (a)
\includegraphics[width=\linewidth]{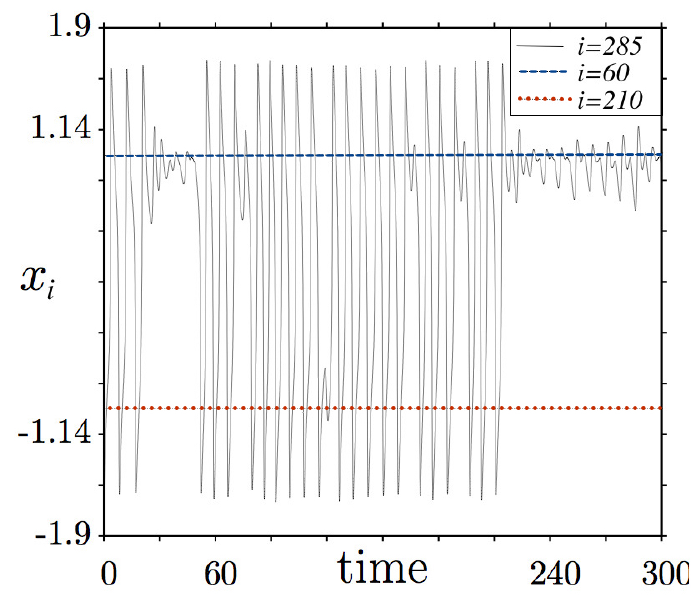}
\center (c)
}
\parbox[c]{.38\linewidth}{
  \includegraphics[width=\linewidth]{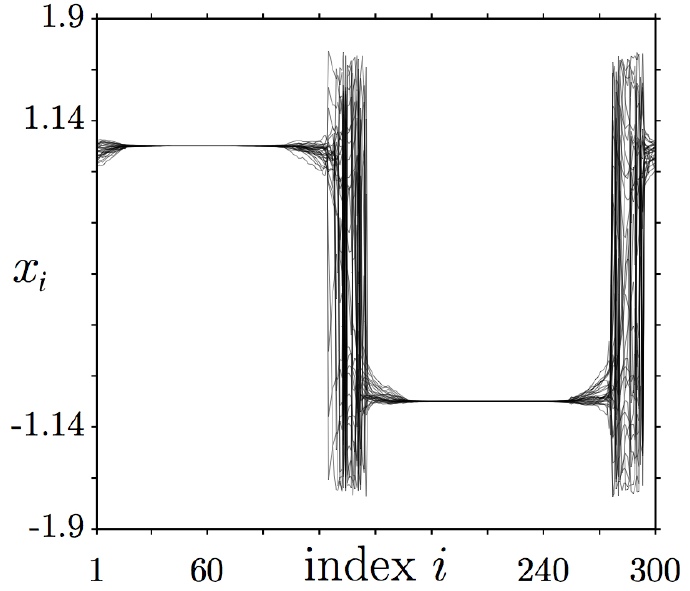}
\center (b)
  \includegraphics[width=\linewidth]{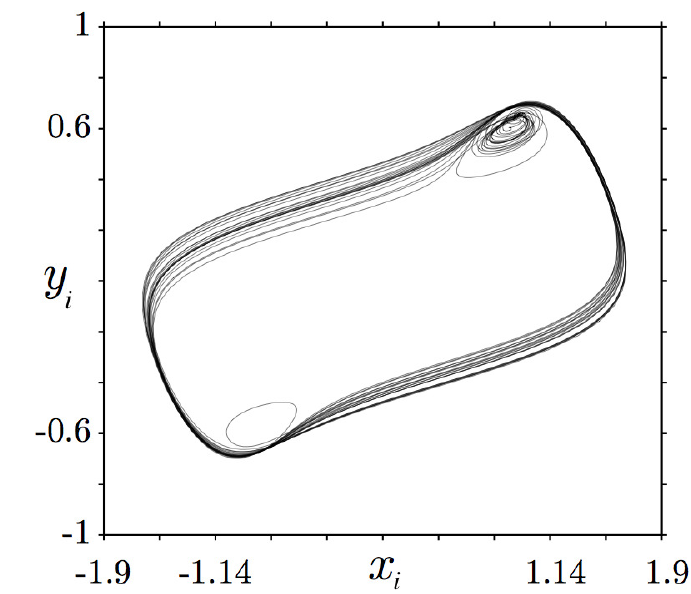}
\center (d)
}
\caption{Chimeras in the bistable ring Eq.\eqref{eq:main}. (a): snapshot of the variable $x_i$, (b): set of 30 snapshots of the variable $x_i$ in different time moments, (c): time realization $x_{i}(t)$ coordinate for the three different selected elements with $i=60;~210;~285$,  (d): the ($x_i,y_i$)-projection of the phase trajectory for the 285th element. Parameters: $r=0.077$, $\sigma=0.34$, $\varepsilon=0.2$, $\gamma=0.7$, $\beta=0.001$, $N$=300}
\label{pic:chimera}
\end{figure}

Such a chimera structure is exemplified in Fig.~\ref{pic:chimera} in more detail. The corresponding space-time plot for this regime is plotted in Fig.~\ref{pic:regimes}(c). 
%
%
The instantaneous spatial profile $x_i$ (Fig.~\ref{pic:chimera}(a)) shows the structure which is typical for the chimera state and consists of two coherence and two incoherence clusters.  
The plot, which has been named as a space-time profile in \cite{Anishchenko-2016}, is shown in Fig.~\ref{pic:chimera}(b). It illustrates a variety of instantaneous spatial profiles (snapshots) at different time moments. Thus, we can observe spatial changes in time. The space-time profile indicates that the oscillators in the center of the coherence cluster centers rest in the equilibrium states and do not oscillate. 
Weak oscillations appear only near the boundary of the coherence cluster. Inside the incoherence cluster, the amplitude of these oscillations increases and switchings between the wells become possible. The irregular character of oscillations in the center of the incoherence clusters and the absence of oscillations in the coherence cluster centers can be seen from the time realization $x_{i}(t)$ plotted in Fig.~\ref{pic:chimera}(c). Oscillations for three different elements of the ring are shown in this graph. The 60th element belongs to the coherence cluster and remains in the right equilibrium point. Its oscillations are given by the dashed line (blue online). The dotted curve (red online) corresponds to the 210th oscillator located in the left equilibrium point. It belongs to the coherence cluster center. The 285th oscillator (black solid line) demonstrates irregular oscillations in time. This element belongs to the incoherence cluster. The phase trajectory projection for this oscillator is pictured in Fig.~\ref{pic:chimera}(d) and also gives evidence for the irregular behavior of the oscillator in time.

The "bistable" chimeras, which have been found in the ring Eq.\eqref{eq:main} for the chosen values of the coupling parameters, represent structurally stable patterns. In contrast to the chimeras which exist in ensembles of chaotic oscillators \cite{Omelchenko-2011, Omelchenko-2012, Anishchenko-2016}, these chimeras have wide basins of attraction. At the same time, the behavior observed in Eq.\eqref{eq:main} depends on the initial conditions and thus, the effect of multistability takes place for any values of $\sigma$ and $r$. 

\begin{figure}[!ht]
\centering
\parbox[c]{.4\linewidth}{
  \includegraphics[width=\linewidth]{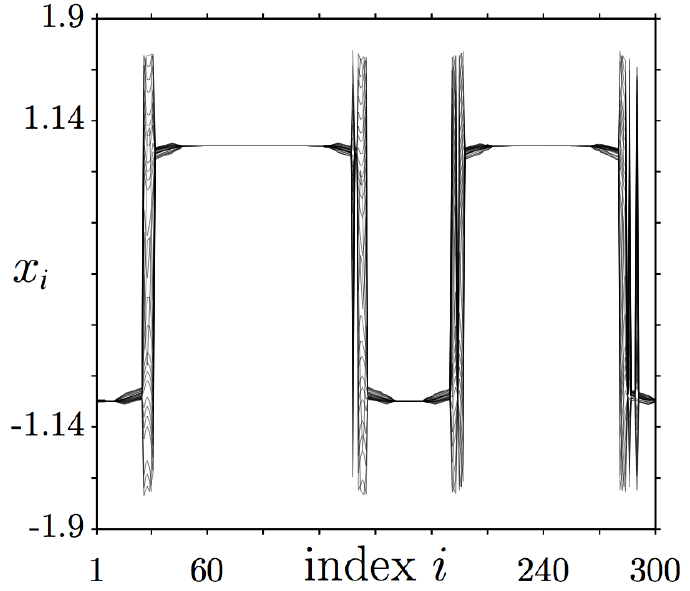}
\center (a)
}
\parbox[c]{.4\linewidth}{
  \includegraphics[width=\linewidth]{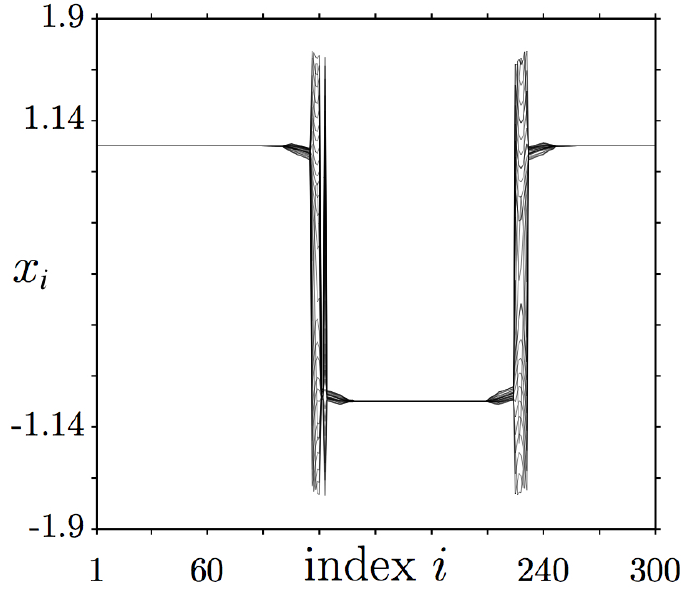}
\center (b)
}
\caption{Multistability of chimera structures in the bistable ring Eq.\eqref{eq:main}. set of 30 snapshots of the variable $x_i$ in different time moments for two chimeras observed for different initial conditions. Parameters: $r=0.05$, $\sigma=0.19$, $\varepsilon=0.2$, $\gamma=0.7$, $\beta=0.001$, $N$=300}
\label{pic:multistability}
\end{figure}

So, if for any values of $\sigma$ and $r$, all the oscillators at the initial time moment are located in the basin of attraction of only one of the equilibria, then the resulting regime represents the corresponding spatio-uniform equilibrium state. For a relatively small value of $r$, the multistability of chimera structures is typical for the ring Eq.\eqref{eq:main}, i.e., chimeras with different numbers of incoherence clusters ("multihead" chimeras) can be observed for different initial conditions. Examples of two coexisting chimeras ("two-head" and "four-head") are shown in Fig.~\ref{pic:multistability}. If $r$ continues decreasing, then the number of coexisting structures increases and the size of incoherence regions decreases tending to one-two oscillators in the limit of local coupling. In this case, the assumption about chimera existence becomes improbable. Only "two-head" chimeras are implemented for different initial conditions with increasing $r$.

%
%
Our studies have shown that the type of chimera structure, which has been revealed in the ring Eq.\eqref{eq:main}, exists in a narrow but finite range of the parameter $\gamma$ and $\beta$ values, which control the FHN oscillator dynamics. The condition for the chimera existence is the bistable regime of the oscillators with two stable foci as well as the closeness to the threshold of the emergence of self-sustained oscillations. 

When decreasing $\gamma$ and moving away from this threshold, the oscillators which form incoherence clusters stop to oscillate and are distributed between the two equilibrium states. The transition to a stationary spatial structure is realized. When moving from the bistability region to the excitable dynamics area with increasing $\beta$, the chimeras also disappear and are replaced by the regime of traveling waves.
 
\section*{Conclusion}

The chimera states which we have revealed in the network of FHN oscillators Eq.\eqref{eq:main} in the bistable regime are different from the previously described class of chimeras. This is a new type of chimera structures which cannot be observed in systems consisting of elements which are similar to phase oscillators, harmonic generators or chaotic oscillators (for example, the logistic map or the R\"{o}ssler oscillator). 
On the one hand, these chimeras are similar to the structures which appear as a result of dynamically sustaided bistability in the ensemble of harmonic self-sustained oscillators (amplitude chimera and chimera death \cite{Zakharova-2014}). However, on the other hand, they have significant differences. 
The new chimeras in Eq.\eqref{eq:main} have "motionless" clusters of oscillators which are in equilibrium states, but, unlike amplitude chimeras, they are not completely statical as in the case of chimera death. 
They also differ significantly from the amplitude and phase chimeras in ensembles of chaotic oscillators \cite{Omelchenko-2011, Omelchenko-2012, Anishchenko-2016} because for them, developed irregular oscillations are observed only within incoherence regions. Besides, the chimeras under study have essentially stronger structural stability than the chimeras in \cite{Omelchenko-2011, Omelchenko-2012, Anishchenko-2016}.

The reason of occurrence and the peculiarities of the described chimera structures in the ring Eq.\eqref{eq:main} are apparently related to the bistable character of the network elements. Perhaps, the important role may also play the closeness to the bifurcations of separatrix loop formation, which are observed in a single isolated FHN oscillator for small values of $\beta$ near the threshold of self-sustained oscillations.
Explanation of the mechanisms of emergence of the described chimera structures requires a more detailed study.

\section*{Acknowledgments}
This research is supported by the Russian Science Foundation (grant \#~16-12-10175) and is partly supported in the framework of SFB910.


\clearpage

\section*{References}

\begin{thebibliography}{99}



\bibitem{Panaggio-2015}
M.J. Panaggio, D.M. Abrams, Nonlinearity 28 (2015) R67.


\bibitem{Kuramoto-2002}
Y. Kuramoto, D. Battogtokh, Nonl. Phenom. Complex Syst. 4 (2002) 380--385. 

\bibitem{Abrams-2004}
D.M. Abrams, S.H. Strogatz, Phys. Rev.Lett. 93 (2004) 174102. 

\bibitem{Abrams-2006}
D.M. Abrams and S.H. Strogatz, Int. J. of Bif. Chaos. 16 (2006) 21--37.

\bibitem{Tinsley-2012}
M.R. Tinsley, S. Nkomo, K. Showalter, Nature Phys. 8 (2012) 662-665.

\bibitem{Omel'chenko-2012}
O.E. Omel'chenko, M. Wolfrum, S. Yanchuk, Y.L. Maistrenko, O. Sudakov, Phys. Rev. E. 85 (2012) 036210.

\bibitem{Maistrenko-2015}
Y. Maistrenko, O. Sudakov, O. Osir, V. Maistrenko, NJP. 17 (2015) 073037.


\bibitem{Zakharova-2014}
A. Zakharova, M. Kapeller, E. Sch\"{o}ll, Phys. Rev. Lett. 112 (2014) 154101.

\bibitem{Kapitaniak-2014}
T. Kapitaniak, P. Kuzma1, J. Wojewoda1, K. Czolczynski, Y. Maistrenko, Scientific Reports. 4 (2014) 6379.

\bibitem{Omelchenko-2015a}
I. Omel'chenko, A. Zakharova, P. H\"{o}vel, J. Siebert, E. Sch\"{o}ll, Chaos. 25 (2015) 083104.

\bibitem{Vullings-2016}
A. V\"{u}llings, J. Hizanidis, I. Omel'chenko, P. H\"{o}vel, NJP. 16 (2016) 123039.



\bibitem{Omelchenko-2011}
I. Omel'chenko, Y. Maistrenko, P.H\"ovel, E. Sch\"{o}ll, Phys. Rev.Lett. 106 (2011) 234102.

\bibitem{Omelchenko-2012}
I. Omel'chenko, B. Riemenschneider, P. H\"{o}vel, Y. Maistrenko, E.Sch\"{o}ll, Phys. Rev. E.85 (2012) 026212.

\bibitem{Semenova-2015}
N. Semenova, A. Zakharova, E. Sch\"{o}ll, and V. Anishchenko, Europhys.~Lett. 112 (2015) 40002.

\bibitem{Anishchenko-2016}
S.A. Bogomolov, A.V. Slepnev, G.I. Strelkova, E. Sch\"{o}ll, V.S. Anishchenko, Comm. in Nonl. Sci. and Numer. Sim. 43 (2017) 25-36.

\bibitem{Vadivasova-2016}
T.E. Vadivasova, G.I. Strelkova, S.A. Bogomolov, V.S. Anishchenko, Chaos. 26:9 (2016) 093108. 

 
 \bibitem{Omelchenko-2013}
I. Omel'chenko, O.E. Omel'chenko, P. H\"{o}vel, E. Sch\"{o}ll, Phys. Rev. Lett. 110 (2013) 224101.
 
\bibitem{Omelchenko-2015b}
I. Omel'chenko, A. Provata, J. Hizanidis, E. Sch\"{o}ll, P. H\"{o}vel, Phys. Rev. E. 91 (2015) 022917.

\bibitem{Semenova-2016}
N. Semenova, A. Zakharova, V. Anishchenko, E. Sch\"{o}ll, Phys. Rev. Lett. 117 (2016) 014102. 


\bibitem{FitzHugh-1961}
R. FitzHugh, Biophysical Journal. 1:6 (1961) 445-466.

\bibitem{Nagumo-1962}
J.S. Nagumo, S. Arimoto, S. Yoshizawa, Proc. of the Institute of Radio Engineers. 50 (1962) 2061-2071.


\bibitem{Ermentrout-2007}
B. Ermentrout, D. Pinto, SIAM News. 40 (2007) 2.

\bibitem{Gorelova-1983}
N.A. Gorelova, J. Bures, Journal of Neurobiology. 14:5 (1983) 353-363.

\bibitem{Huang-2004}
X. Huang, W.C. Troy, Q. Yang et al., J. of Neurosci. 24 (2004) 9897-9902.

\bibitem{Lancaster-2010}
J.L. Lancaster, E.H. Hellen, E.M. Leise, Amer. J. of Physics. 78:1 (2010) 56-63.


\bibitem{Shepelev-2016}
I. A. Shepelev , A. V. Slepnev, T. E. Vadivasova, Comm. in Nonl. Sci. and Numer. Sim. 38 (2016) 206-217. 

\bibitem{Shepelev-2015}
I. A. Shepelev , T. E. Vadivasova, D. E. Postnov, Proc. of~ SPIE. 9448 (2015) 94481V-1.


\bibitem{Nekorkin-2002}
V.I. Nekorkin, M.G. Velarde, Synergetic Phenomena in Active Lattices, Springer, Berlin, 2002.


\end{thebibliography}
\end{document}